\documentstyle[psfig,amssymb,referee]{mn}
  
\newcommand{\etal}{{\it et~al.~}}

\title[Star formation in faint radio sources]
{The 1.4~GHz and H$\alpha$ 
Luminosity Functions and Star Formation Rates from Faint Radio Galaxies}

\author[B. Mobasher \etal ] 
{B. Mobasher$^1$, L. Cram$^2$, A. Georgakakis$^1$ and A. Hopkins$^2$\\
$^1$Astrophysics Group, Blackett Laboratory, Imperial College, 
Prince Consort Rd, London SW7 2BZ, UK\\
$^2$ School of Physics, University of Sydney, Sydney NSW 2006, Australia\\ 
}
\begin{document} 
\maketitle
\begin{abstract} 

  A sample of over 1000 objects selected from a 1.4~GHz survey made by the
  Australia Telescope Compact Array (ATCA) is used to study the properties
  of the faint radio source population. The sample, covering an area of
  $\approx 3$ deg$^2$, is $50\%$ complete to 0.2~mJy. Over 50$\%$ of the
  radio sources are found to have optical counterparts brighter then $R
  \approx 21.5$. Spectroscopic observations of 249 optically identified
  radio sources have been made, using the 2-degree Field (2dF) facility at
  the Anglo-Australian Telescope (AAT). Redshifts and equivalent widths of
  several spectral features (e.g., H$\alpha$ and [OII]$\lambda3727$)
  sensitive to star formation have been measured. On the basis of the
  photometric and spectroscopic data, the optically identified radio
  sources are classified as (i) absorption-line galaxies, (ii) star-forming
  galaxies and (iii) Seyfert-like galaxies.

  The spectroscopic sample is corrected for incompleteness and used to
  estimate the 1.4~GHz and H$\alpha$ luminosity functions (LFs) and
  luminosity density distributions. The 1.4~GHz LF of the star-forming
  population has a much steeper faint-end slope (1.85) than the ellipticals
  (1.35). This implies an increasing preponderance of star-forming galaxies
  among the optically identified (i.e., $z \lesssim 1$) radio sources at
  fainter flux densities. The H$\alpha$ LF of the faint radio population
  agrees with published H$\alpha$ LFs derived from local samples selected
  by H$\alpha$ emission.  This suggests that the star-forming faint radio
  population is coincident with the H$\alpha$ selected population.

  The 1.4~GHz and H$\alpha$ luminosity densities have been used to estimate
  the star formation rates (SFRs). The two SFRs agree, both giving a SFR
  density of 0.032 M$_\odot$ yr$^{-1}$ Mpc$^{-3}$ in the range $z \lesssim
  1$.  Radio selection appears to be as effective as H$\alpha$ selection in
  finding the galaxies which dominate star formation at a given epoch.
  Although the sample contains many galaxies lying beyond $z \approx 0.3$,
  it does not reveal a significant rise in the global star formation rate
  with increasing redshift. This result suggests that the optical
  counterparts of galaxies undergoing vigorous star formation at redshifts
  beyond $z \approx 0.3$ are generally fainter than $R \approx 21$.

\end{abstract} 


\section{Introduction}

The population of apparently bright radio sources ($S_{1.4} \gtrsim
100$~mJy) is dominated by {\it classical radio galaxies}, whose optical
hosts, when detectable, are usually identified as luminous elliptical
galaxies with red colours and absorption-line optical spectra, although
some are hosted by QSOs.  The elliptical hosts sometimes emit nebular
emission lines, which can be quite strong for the most luminous radio
sources \cite{Hin:79}. The lines appear to be excited by an active galactic
nucleus (AGN) radiating a non-stellar continuum \cite{Bau:89}.  The
normalised source count distribution of the classical radio galaxies falls
sharply with decreasing flux density, as a result of cosmological dimming
and strong source evolution for $z \lesssim 2$ \cite{Wal:97}.

The apparently bright radio source population contains, in addition to
these classical radio galaxies, a small proportion of spiral/irregular
galaxies. These are markedly different from the classical radio galaxies.
For example, even samples selected to have large radio-to-optical
luminosity ratios have a radio luminosity smaller than that of most
classical radio galaxies \cite{Con:82}.  The core/jet/lobe radio morphology
of classical radio galaxies is absent, and replaced by centrally
concentrated but nevertheless diffuse emission.  Moreover, a high
proportion of radio-bright spirals have peculiar optical morphologies and
strong emission lines excited by stellar photospheric radiation, apparently
reflecting an enhanced rate of star formation \cite{Con:82}.

Spiral/irregular galaxies do not contribute in significant numbers to radio
source counts for $S_{1.4} \gtrsim 10$~mJy.  However, radio surveys
reaching fainter flux density limits, $S_{1.4} \lesssim 1$~mJy, reveal a
significant change in the slope of the normalised source count distribution
at $S_{1.4} \approx 5$~mJy [Windhort, van Heerde \& Katgert
\shortcite{Win:84}; Condon \shortcite{Con:84}; Fomalont et al.\ 
\shortcite{Fom:84}; Mitchell \& Condon \shortcite{Mit:85}; Windhorst et
al.\ \shortcite{Win:85}].  The change suggests the possible existence of a
numerous population of apparently faint radio sources unlike the classical
radio galaxies.  Optical photometry and spectroscopy of the population
reveals that many of the faint radio sources are associated with blue
galaxies often exhibiting peculiar (compact, interacting and merging)
morphologies. Kron, Koo \& Windhorst \shortcite{Kro:85} and Thuan \& Condon
\shortcite{Thu:87} have shown that these galaxies are quite unlike the
hosts of classical radio galaxies, but are reminiscent of radio-bright
local spiral galaxies.

The central astrophysical importance of the faint radio population lies in
the fact that the unevolving, local luminosity function of the non-AGN
radio source population can explain neither the large number counts nor the
source-count distribution at sub-mJy flux densities \cite{Con:84}.
Attempts to resolve this problem show that significant evolution may have
occurred over the redshift range spanned by the observed population.  For
example, Danese et al.  \shortcite{Dan:87} established consistency by
adopting a model which includes luminosity evolution of a
starburst/interacting population over an $e$-folding time of about 25\% of
the Hubble time.  There are hints that the IRAS galaxy population undergoes
similar evolution, although the IRAS surveys did not extend to the faint
levels corresponding to deep radio surveys \cite{Lon:90}.  Population
evolution of FIR-selected galaxies has been confirmed by deeper FIR
observations made by the {\it Infrared Space Observatory (ISO)}
\cite{Row:97}.  Additional evidence, consistent with the hypothesis of
recent ($z \lesssim 1$) rapid evolution of a large fraction of the faint
radio population, has come from the improved redshift distribution
statistics for the optical counterparts of faint radio sources reported by
Benn et al.\ \shortcite{Ben:93} and Hopkins et al.\ \shortcite{Hop:99}.

The cosmological significance of this evolving population has been
clarified by optical/UV studies using deep ground-based observations and
data from the {\it Hubble Space Telescope (HST)}. Studies by Madau et al.\ 
\shortcite{Mad:96} imply not only that the space density of star-forming,
morphologically disturbed galaxies increases with redshift to $z \gtrsim
1$, but also that this increase is associated with a corresponding increase
in the global star formation rate in the Universe. A similar picture
emerges from observations of faint sub-mm sources \cite{Bla:98}.  

The 1.4~GHz luminosity of a spiral/irregular galaxy can be used to estimate
its star formation rate ($SFR$), assuming the appropriate calibration
(Condon \& Yin 1990; Condon 1992). Using this result and published 1.4~GHz
luminosity functions, Cram \shortcite{Cra:98a} has argued that the global
rate of star formation appears to be dominated by galaxies with $SFR
\gtrsim 15$ M$_\odot$ yr$^{-1}$. This implies that galaxies with $P_{1.4}
\gtrsim 5 \times 10^{22}$ W Hz$^{-1}$ are likely to dominate the global
$SFR$. Such a galaxy would be brighter than 10 $\mu$Jy at $z = 1$, implying
that deep radio surveys probe the population which dominates global star
formation to cosmologically significant redshifts.

The discoveries summarised above suggest that the faint radio source
population may well contain {\it all} of the galaxies which dominate the
global star-formation processes revealed by optical/UV diagnostics, at
least for redshifts out to $z \approx 1$ at the current limits of radio
surveys.  However, the implications of this have been explored directly for
only a few dozen objects [Benn et al. \shortcite{Ben:93}; Hammer et al.
\shortcite{Ham:95}].  In the present study, we use a sample of over 420
faint, radio-selected galaxies with optical data to define a well
characterised sub-sample of star-forming objects, many (over $50\%$) with
measured redshifts. From this sub-sample, we use the radio and H$\alpha$
luminosities to estimate luminosity functions, luminosity densities and
star-formation rates in the era $z \lesssim 1$.  We show that the radio and
optical properties of this sample are consistent with one another, and with
the properties of samples from the {\it local} Universe.  We use the
results to provide an unbiased estimate of the star-formation history (free
from dust effects) in the surveyed volume.

In the following section we summarise the observations. Section 3
presents the luminosity functions and star formation rates, and
Section 4 discusses the results from this investigation.

\section{Observations}

The observations have been accumulated during several campaigns connected
with the Phoenix Deep Survey.  The criteria used to select the field, as
well as the radio observations and data reduction methods, are discussed by
Hopkins et al.\ \shortcite{Hop:98}.  The radio observations were made with
the Australia Telescope Compact Array (ATCA) at 1.4~GHz, covering an area
of diameter $2^\circ$, centered at ${\mathrm RA}(2000)=01^{\mathrm
  h}~14^{\mathrm m}~12\fs16$ ; ${\mathrm
  Dec.}(2000)=-45^{\circ}~44'~8\farcs0$.  A total of 1079 radio sources are
detected to $S_{1.4} \approx 0.1$~mJy.  The radio flux distribution for the
entire survey is presented in Fig. 1 (solid line).  The survey is 60\%
complete to $S_{1.4} \approx 0.3$~mJy, with a total of 804 sources detected
to this flux limit.  The completeness drops to $50\%$ at $S_{1.4} \approx
0.2$~mJy, with a total of 964 galaxies to this limit. The $80\%$ 
completeness limit is achieved at about $S_{1.4}=0.4$ mJy with 656 galaxies
detected. In the present study, we consider galaxies with $S_{1.4} > 0.2$ mJy
and refer to this in the following sections as the `complete' radio sample.

CCD observations of the entire radio survey area were attempted in the $V$
and $R$-bands, using the Anglo-Australian Telescope (AAT). The observations
were performed to a magnitude limit of $R=22.5$, corresponding to the limit
expected for successful spectroscopy with the two-degree field facility
(2dF) of the AAT. Over $50\%$ of the radio sources 
have optical identifications, corresponding to a total of 420
galaxies to $S_{1.4} \approx 0.3$~mJy and 490 galaxies to 0.2~mJy.  The
radio flux distribution for galaxies with optical IDs is shown in Fig. 1
(dotted line). The optical photometry and radio-optical source
identifications are described in Georgakakis et al.\ \shortcite{Geo:98}.

\begin{figure} 
\centerline{\psfig{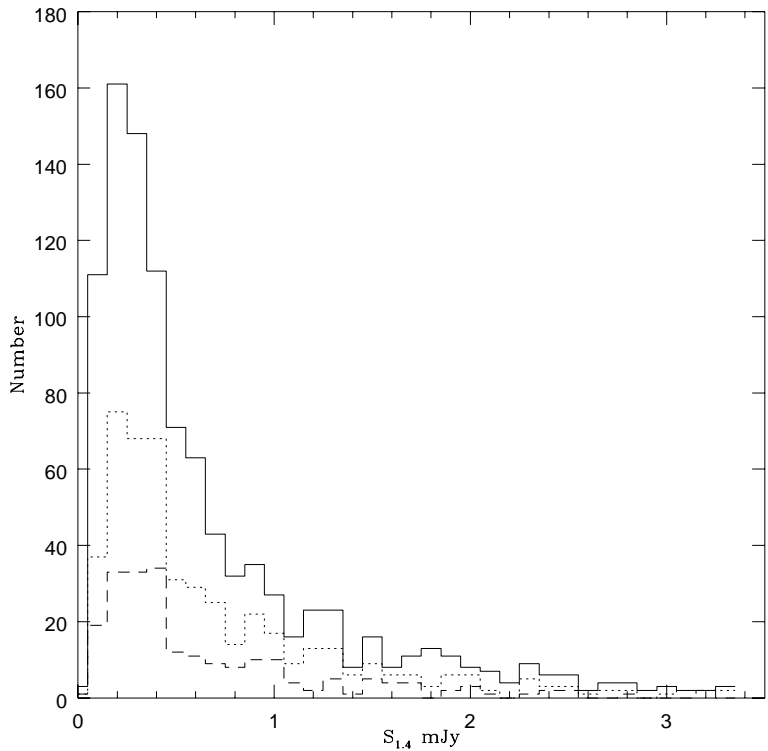}}
\caption{1.4~GHz radio flux distribution of detected radio sources
  (solid line), radio sources with optical identifications (dotted line)
  and radio sources with spectroscopic measurements (dashed line).}
\label{diagnostic}
\end{figure}

Spectroscopic observations were carried out for a random sample of the
optically identified radio galaxies to a limiting magnitude of $R=21.5$,
using the 2dF facility. Accurate redshifts were successfully measured for a
total of 228 radio sources, using several lines, including H$\beta$,
H$\alpha$, [OII]$\lambda3727$ and [OIII]$\lambda5003$ \cite{Geo:98}.  Of
this number, 210 galaxies are in the `complete' radio survey (i.e., have
$S_{1.4} > 0.2 $~mJy), implying $\approx 50\%$ completeness for the
spectroscopic sample used here.  The fraction of galaxies with
spectroscopic data at any given radio flux is presented in Fig. 1 (dashed
line).  Diagnostic spectral line ratios were estimated from these spectra
and used to classify the galaxies in the spectroscopic sample.  This is
explained in detail in Georgakakis et al. (1998) where the galaxies are
classified as (i) elliptical, (ii) starforming, (iii) Seyfert and (iv) a
small number of objects for which the spectral quality was not high enough
to allow an accurate classification. Examples of the 2dF spectra of the
radio sources classified to the above four sub-types are shown in Fig. 2.

\begin{figure} 
\centerline{\psfig{figure=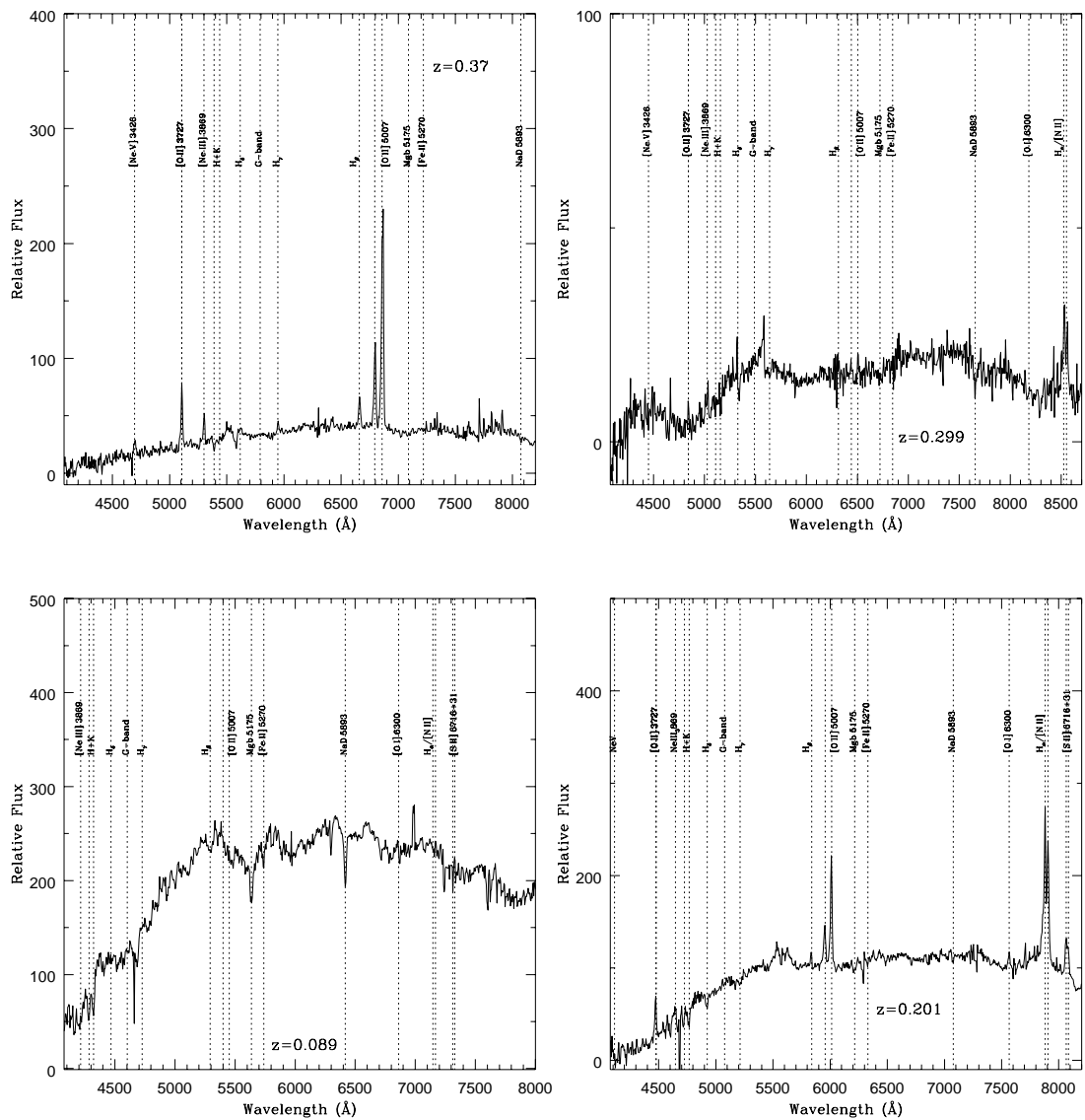,width=1\textwidth,angle=0}}
\caption{2dF spectra for the 4 types of galaxies classified here. 
The identified spectral lines are marked with the estimated redshift of the 
objects specified. The spectra correspond to; Elliptical (bottom left), 
Star-forming (top left), Seyfert 2 (bottom right) 
and `unclassified' (top right)}
\end{figure}

Spectroscopic signatures of probable star formation (mainly
[OII]$\lambda3727$ and H$\alpha$) are detected in the spectra of $63\%$ of
the faint radio sources.  The redshift distributions for the complete radio
sample ($S_{1.4} > 0.2$~mJy) is compared in Fig. 3 with that of the
sub-sample showing evidence of star formation activities as revealed from
their spectra.  The majority of star-forming radio sources brighter than
the optical magnitude limit of this survey have $z \lesssim 0.3$.

\begin{figure} 
\centerline{\psfig{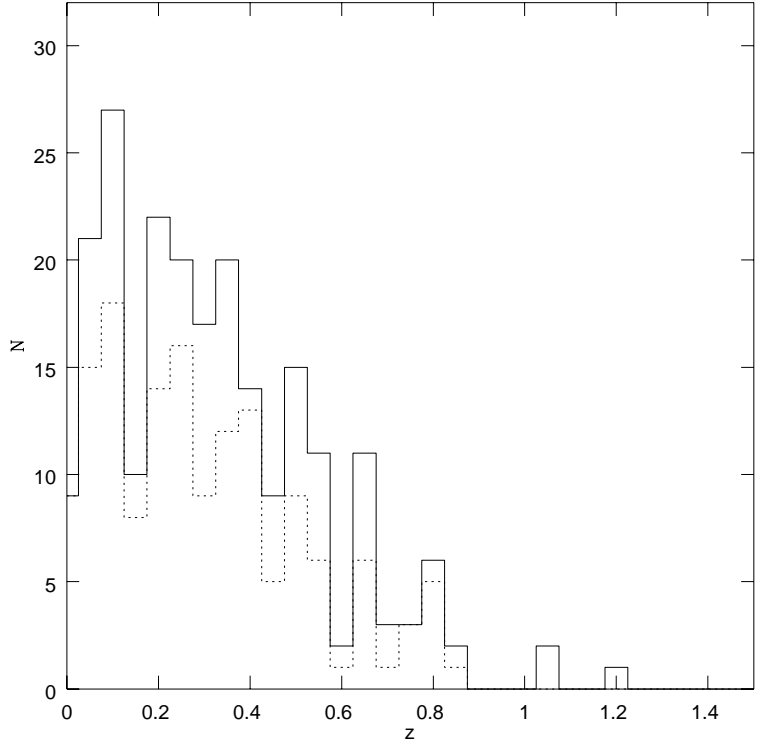}}
\caption{
  The redshift distribution (solid line) of the Phoenix sample compared
  with the distribution of galaxies with H$\alpha$ (or [OII]$\lambda$3727)
  detections (dotted line).  }
\label{diagnostic}
\end{figure}
 
One aim of this study is to derive quantitative optical measures of the
star formation rate, using H$\alpha$ wherever possible, to compare with the
results from the radio data.  However, an absolute flux calibration is not
available for our 2dF spectra.  Thus, the H$\alpha$ luminosity
($P_{H\alpha}$) of each galaxy has been estimated from its measured
H$\alpha$ equivalent width, $EW_{\alpha}$, and the $R$-band magnitude.  The
spectral synthesis models of Bruzual and Charlot \shortcite{Bru:93} were
then used to estimate the $R$-band $K$-corrections and the rest-frame
continuum luminosity at H$\alpha$, $C_{\alpha}$ \cite{Geo:98}.  The
H$\alpha$ luminosity is then estimated using

\[ 
P_{H\alpha} = C_{\alpha}\times EW_{H_\alpha}.
\]
For the few galaxies with $z > 0.3$, H$\alpha$ shifts out of the observed
spectral range. For these, we used [OII]$\lambda3727$ as a proxy for
H$\alpha$, converting its equivalent width using the relation \cite{Ken:92}
\[
 EW_{OII} = 0.4 \times EW_{H\alpha}.
\]
As noted by Kennicutt \shortcite{Ken:92}, the scatter about this
correlation has an RMS dispersion of about 50\%, due in part to
galaxy-to-galaxy differences in extinction.

\section{Luminosity functions and star formation rates}

In this section we estimate the 1.4~GHz and H$\alpha$ luminosity functions
for our sample. These will then be used to determine the luminosity density
distribution and, for the star-forming galaxies, the star formation rate.
To ensure an adequate sample size, we first assume that the sample
statistics are independent of redshift (i.e. that there is no dependence of
the selection functions on redshift). We consider two radio samples,
consisting of the elliptical and the star-forming populations, as
classified from their optical spectra (the small proportion of Seyfert or
unclassified galaxies is excluded). Our H$\alpha$ sample combines all
galaxies in the radio-selected sample which exhibit H$\alpha$ emission,
since this would appear to provide a sample selected in much the same way
as the objective prism sample of Gallego et al.  \shortcite{Gal:95}.

\subsection{Sample completeness and selection effects}

There are two sources of incompleteness in the present spectroscopic
survey.  Firstly, only radio sources with optical counterparts have been
observed spectroscopically. This biases the spectroscopic sample against
optically faint galaxies (e.g., distant, dusty and/or low surface
brightness objects).  Secondly, only a randomly selected sub-sample of the
optically identified galaxies which satisfy the optical magnitude range
accessible by the 2dF have been observed spectroscopically.

To correct for the first bias we define a completeness function at any
given radio flux density, $\eta_1~(S_{1.4})$, as the ratio of the number of
radio sources with optical identifications to the total number of sources
in the complete radio survey.  This completeness function is displayed in
Fig. 4a. The second bias can be corrected using a completeness function
at any given optical magnitude, $\eta_2 (R)$, defined as the ratio of the
number of galaxies with measured spectra, to the total number of sources
with optical counterparts in the `complete' radio survey ($S_{1.4} >
0.2$~mJy).  This completeness function is shown in Fig. 4b. The final
completeness function is then $p_1 (S_{1.4}, R) = \eta_1 (S_{1.4}) \ \eta_2
(R)$.

\begin{figure} 
\centerline{\psfig{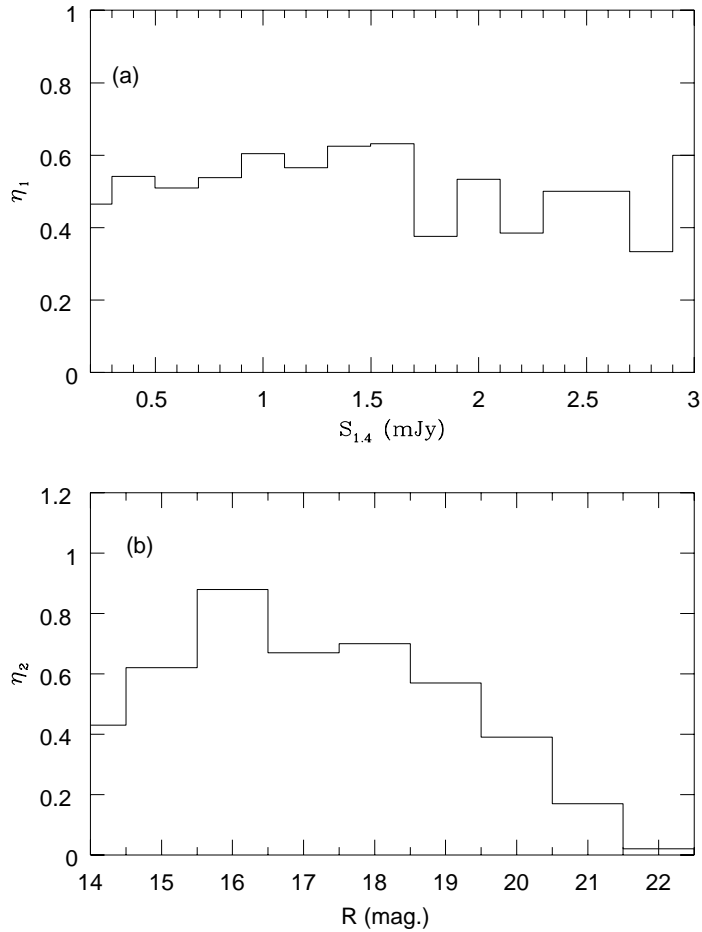}}
\caption{  
  (a) The selection function $\eta_1 (S_{1.4})$ defined as the ratio of the
  number of faint radio sources with optical identifications to the total
  number of radio sources in the complete radio survey.  (b) The selection
  function $\eta_2 (R)$ defined as the ratio of the number of galaxies with
  measured spectra to the total number of sources with optical counterparts
  in the complete radio survey.}
\label{Fig4}
\end{figure}

The H$\alpha$ sample systematically excludes optically identified but faint
galaxies which yield spectra with low S/N ratios and hence no reliable
detection of H$\alpha$ (or [OII]$\lambda$3727).  A completeness function
to correct for this selection effect, $p_2 (R)$, will depend on the optical
magnitude of an identification, and is defined as the ratio of the number
of objects with H$\alpha$ detection to the total number of sources with
optical identifications in the complete radio survey.  The function $p_2
(R)$ is shown in Fig. 4c.

\begin{figure} 
\addtocounter{figure}{-1}
\centerline{\psfig{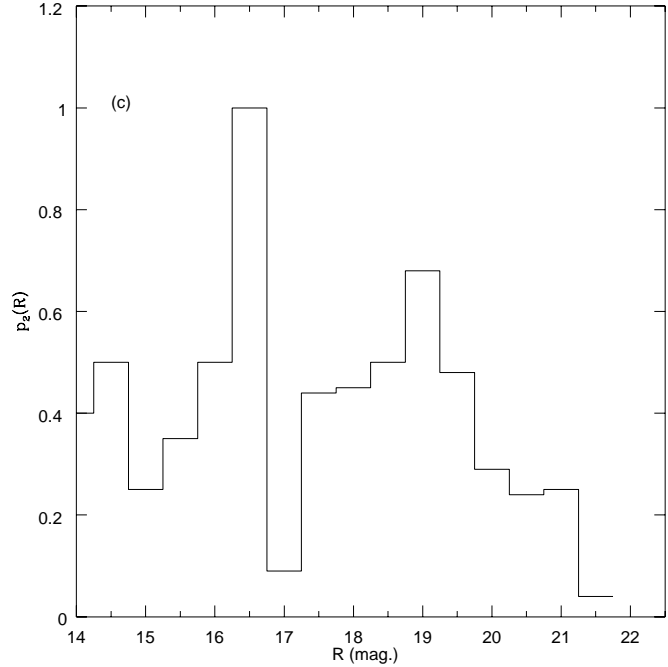}}
\caption{ 
  (c) The selection function of the H$\alpha$ survey, $p_2 (R)$, defined
  as the ratio of the number of galaxies with H$\alpha$ detected to the
  total number of galaxies with optical identification.}
\label{Fig4}
\end{figure}
 
\subsection{The 1.4~GHz and H$\alpha$ luminosity functions}

Using the completeness functions derived above, we correct the sample for
incompleteness and selection effects and estimate its radio [$\Phi_{1.4}
(P_{1.4})$] and H$\alpha$ [$\Phi_{\alpha}(P_{H\alpha})$] luminosity
functions as
\[
 \Phi_k(P_k) = \Sigma_i ({1\over V_{max,i}}),
\]
where the summation is over the number of galaxies in radio or H$\alpha$
flux bins in the complete sample, with $V_{max}$ given by
\[
 V_{max} = {\Omega\over 4\pi} {c\over H_0} \int_{z_{1}}^{z_{2}} {p_k\ 
  {d_L}^2\over (1+z)^2 (1+2 q_0 z)^{1/2}} dz.
\]
Here, $p_k$ is the completeness function corresponding to the radio ($k=1$)
and H$_\alpha$ ($k=2$) surveys, $d_L$ is the luminosity distance, and the
integral limits are $z_1 = max(z_{rad},z_{opt},z_{min})$ and $z_2 =
min(z'_{rad},z'_{opt},z_{max})$. The quantities $z_{rad}$ and $z_{opt}$
are, respectively, the redshifts corresponding to the bright radio flux and
optical $R$-band magnitude limits of the survey while $ z'_{rad}$ and
$z'_{opt}$ are the maximum redshifts at which a source could be included
(at the flux limit of the survey) in the radio sample and be detected at
the optical wavelength. The quantities $z_{min}$ and $z_{max}$ are
respectively the minimum and maximum redshifts covered by galaxies in the
radio survey.  $K$-corrections are estimated in the $R$-band using model
spectral energy distributions for intermediate-type spirals (section 2),
and at 1.4~GHz assuming an spectral energy distribution proportional to
$\nu^{-0.7}$.

A parametric form which successfully models the luminosity functions of
the IRAS \cite{Sau:90} and faint radio \cite{Row:93} sources has been adopted
to fit the luminosity functions in this study. It is the function
\[
\Phi (P)\ d (log\ P) 
= C^\ast ({P\over P^\ast})^{1-\alpha}\ \exp [-{1\over 2\sigma^2}
log_{10}^2 (1 + {P\over P^\ast})]\ d (log\ P),
\]
where $C^\ast$, $P^\ast$, $\sigma$ and $\alpha$ are the fitted parameters.
Fig. 5 displays the radio (1.4~GHz) LF for all the galaxies in the
`complete' radio survey (Fig. 5a), the star-forming population (Fig.
5b) and the ellipticals (Fig. 5c). The error bars represent uncertainties
assuming Poisson statistics. 
The $<V/V_{max}>$ estimates are given in Table 1. The parametric fits to 
the radio LFs, assuming the above form, 
are carried out for cases with no evolution and assuming luminosity evolution 
in the form $P_{1.4} (z) = P_{1.4} (0)\ (1+z)^{3}$. This form of 
luminosity evolution is adopted to be consistent with recent optical
(Lilly et al 1996), far-infrared (Kawara et al 1998) and radio
(Rowan-Robinson et al 1993) observations. The fits were performed 
to the LFs corresponding to all the galaxies in the `complete' radio survey, 
the starburst population and elliptical galaxies. 
Figure 5 compares the results for the two cases, assuming no-evolution 
(solid line) 
and luminosity evolution (dotted line). The estimated 1.4 GHz LF parameters
are listed in Table 1. 
For a given population of galaxies in Figures 5a, 5b and 5c,  
we find no significant difference in the radio LFs between the two cases
(assuming no evolution or luminosity evolution in the form $(1+z)^3$). This
result argues against the presence of significant luminosity 
evolution in the radio LFs in the redshift range, 
$0 < z < 1$, covered by the present survey. 
 
\begin{figure} 
\centerline{\psfig{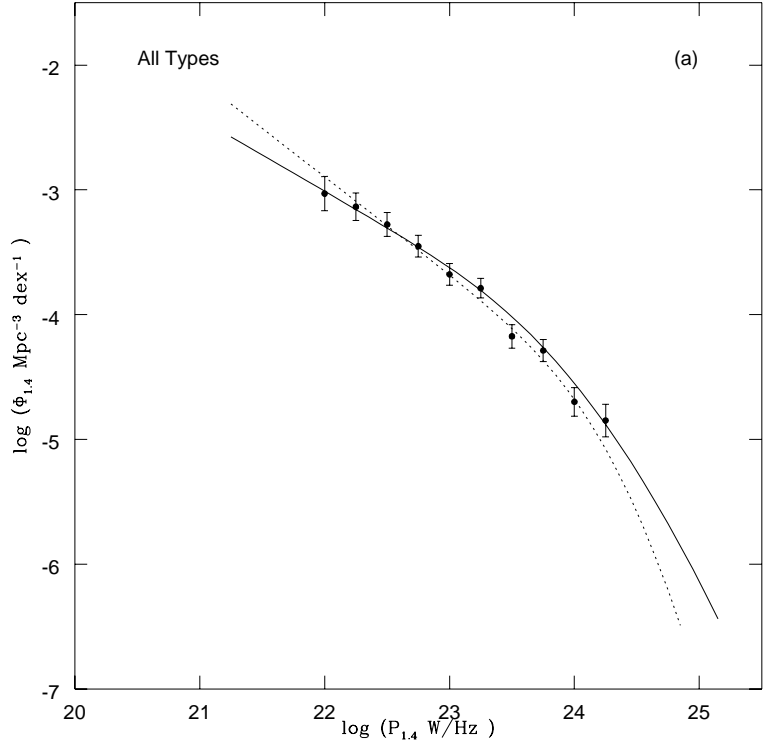}}
\caption{ 
  (a) The 1.4~GHz luminosity functions for all galaxies in the `complete'
  radio survey (filled circles). The sample is corrected for
  incompleteness. The curve is a parametric fit to the LF, using the
  function described in the text and
  assuming no evolution (solid line)
  and luminosity evolution in the form $(1+z)^3$ (dotted line). 
.  The LF is estimated in the range $z<1$.}
\end{figure}

The LF for the star-forming radio sources in Fig. 5b is compared with
Condon's \shortcite{Con:89} estimates from the bright spirals (open
circles), starbursts (filled triangles) and a sample of UGC galaxies 
with diameter $> 1'$ (asterisks). 
The present sample has a mean redshift of $<\!z\!> =0.3$, compared to 
$<\!z\!> =0.05$ for galaxies from Condon (1989) and Sadler et al (1989). 
Therefore, the comparison between these samples
in Figures 5b and 5c assumes no evolution for the radio sources 
in the range $0 < z < 0.3$, as justified from the above results. 
The independently derived LFs show 
excellent agreement in both
shape and normalisation. Moreover, extrapolation of our radio LF in Fig.
5b to faint fluxes ($P_{1.4} < 10^{21}$ W/Hz) agrees well with the
faint-end of the radio (1.4~GHz) LF for the spiral and irregular galaxies
in Condon \shortcite{Con:89}.  Fig. 5c compares the radio LF derived here
with that of the elliptical galaxy radio luminosity function from Sadler,
Jenkins \& Kotanyi \shortcite{Sad:89} converted to 1.4~GHz assuming
$S_{\nu} \propto \nu^{-0.7}$. Considering that these samples have different
selection criteria, depths and completeness limits, the agreement between
the two LFs in Fig. 5c is remarkably good.

\begin{figure}
\addtocounter{figure}{-1}
\centerline{\psfig{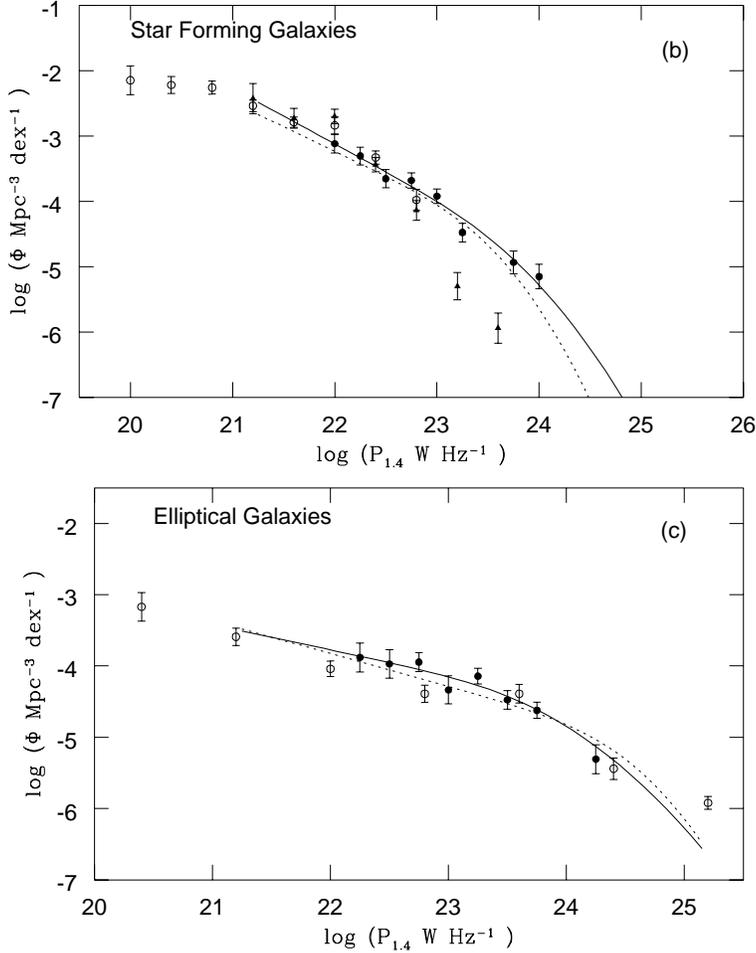}}
\caption{ The 1.4~GHz luminosity functions for the faint radio population,
  corrected for incompleteness. Error bars are estimated assuming Poisson
  distribution.  The curves present parametric fits to the LFs, using the
  function described in the text and assuming no evolution (solid lines)
  and luminosity evolution in the form $(1+z)^3$ (dotted lines). 
  The LFs are estimated in the range $z<1$
  and correspond to (b) star-forming galaxies here (filled circles) compared
  to the spiral/star-forming (open circles) and starburst (filled triangles) 
  galaxies
  from Condon (1984) and (c) elliptical galaxies from this study (solid
  circles) compared with ellipticals from Sadler et al. (1990)-(open
  circles).}
\label{Fig5}
\end{figure}

\begin{table*}
\pagestyle{empty}
\caption{Parameters of the analytical fits to the 1.4~GHz 
  luminosity functions and the $<V/V_{max}>$ estimates. The results are
  presented for all the faint radio sources in the present survey and for
  the starforming and elliptical sub-classes.Two different cases, assuming
  no evolution and luminosity evolution in the form $P_{1.4} (z) = P_{1.4} (0)
  (1+z)^3$ are considered.}
\begin{tabular}{lccccccc}
  & $n$ & $ <V/V_{max}> $ & $P^\ast_{1.4}$ & $\alpha$ & $\sigma$ & 
$C^\ast_{1.4}$ \\ 
  & &                   & W Hz$^{-1}$&& & Mpc$^{-3}$\\ 
          &     &                      &      &      &                     \\  
All types & 192 &0.45 & $(9.59^{+2}_{-2})\times 10^{22}$ & 
$1.58^{+0.12}_{-0.08}$ & $0.80^{+0.10}_{-0.10}$ & 
$(2.63^{+0.40}_{-0.50})\times 10^{-4}$\\
          &     &                      &      &      &                     \\
Starforming & 75 &0.45 & $(1.18^{+0.50}_{-0.23})\times 10^{23}$ & 
$1.85^{+0.15}_{-0.10}$ & $0.67^{+0.13}_{-0.15}$ & 
$(9.28^{+2}_{-2})\times 10^{-5}$\\ 
          &     &                      &      &      &                     \\ 
Ellipticals & 63 &0.42 & $(1.16^{+0.35}_{-0.36})\times 10^{23}$ & 
$1.35^{+0.25}_{-0.20}$ & $0.75^{+0.10}_{-0.15}$ & 
$(7.15^{+2}_{-2})\times 10^{-5}$\\ 
          &     &                      &      &      &                     \\
{\bf Luminosity}&{\bf Evolution }      &      &      &                     \\
          &     &                      &      &      &                     \\
  & $n$ & $ <V/V_{max}> $ & $P^\ast_{1.4}$ & $\alpha$ & $\sigma$ & 
$C^\ast_{1.4}$ \\ 
  & &                   & W Hz$^{-1}$&& & Mpc$^{-3}$\\ 
          &     &                      &      &      &                     \\  
All types & 192 &0.45 & $(5^{+1}_{-1})\times 10^{23}$ & 
$1.78^{+0.05}_{-0.10}$ & $0.47^{+0.33}_{-0.10}$ & 
$(6^{+1.5}_{-1})\times 10^{-5}$\\
          &     &                      &      &      &                     \\  
Starforming & 75 &0.45 & $(1.05^{+0.80}_{-0.20})\times 10^{23}$ & 
$1.75^{+0.15}_{-0.10})$ & $0.50^{+0.30}_{-0.05}$ & 
$(1^{+0.80}_{-0.20})\times 10^{-4}$\\ 
          &     &                      &      &      &                     \\ 
Ellipticals & 63 &0.42 & $(1^{+1}_{-0.2})\times 10^{24}$ & 
$1.46^{+0.20}_{-0.05}$ & $0.50^{+0.10}_{-0.20}$ & 
$(1.8^{+1}_{-0.20})\times 10^{-5}$\\ 
   
\end{tabular}
\end{table*}   

The H$\alpha$ LF of the sub-mJy radio sources 
with H$\alpha$ detection is shown in Figure 6. 
The error bars are estimated assuming
Poisson statistics.  Fits to the analytical form discussed above are also
shown, and the resulting parameters are listed in Table 2. The H$\alpha$ LF
agrees with that determined from an H$\alpha$ survey of field galaxies
\cite{Gal:95}.  (The faintest point, at log$(P_{H\alpha})= 33.5$ W,
significantly deviates from the fitted LF in both studies and is therefore
excluded).  Applying a Schechter LF fit to the present sample gives essentially
the same parameters as those derived in \cite{Gal:95}, implying that the 
result here is independent from the parametric form of the LF. 
Considering the selection effects in the present survey
(Section 2) the agreement is reassuring, and implies that
H$\alpha$-emitting objects in the faint radio population are similar to the
normal H$\alpha$-emitting galaxies in the local Universe.

The slope of the faint-end of the radio LF increases from $\alpha=1.35$ for
ellipticals (Fig. 5c) to $\alpha=1.85$ for the star-forming population
(Fig. 5b). A similar result was also reported by Condon \shortcite{Con:89}.
It implies that in the redshift range $0 < z < 1$ and at modest radio
powers ($P_{1.4} \lesssim 10^{23}$ W/Hz), there is an abundance of
star-forming galaxies compared to ellipticals.  At the median redshift of
our survey, $z \approx 0.3$, galaxies with $P_{1.4} \approx 10^{23}$ W
Hz$^{-1}$ have $S_{1.4} \approx 0.2$~mJy, and most of them have optical
counterparts which will have been detected optically at a limit of $R
\approx 22$. Thus, we may conclude that for the optically identified part
of the sub-mJy population, the number of star-forming ``IRAS-type''
galaxies exceeds the number of ellipticals.  Further observations of
samples which are complete to known radio and optical limits would be of
great value in elucidating the connection between the shape of the LF for
$z \lesssim 1$, possible evolutionary effects, and the observed source
count distributions including objects which have not yet been optically
identified.

There are seven detections of elliptical galaxies with H$\alpha$ in the
present survey. It is important to note that sub-mJy radio sources
classified as elliptical galaxies have rest frame $V-R$ colours close to
those of `normal' ellipticals ($V-R \approx 0.6$) and follow their $V-R$
colour-magnitude relation (Georgakakis et al. 1998). Therefore, although the
radio emission may be from an AGN, there is little evidence of AGN emission
in the optical continuum. It is not excluded that both the radio and the
H$\alpha$ emission from these objects is related at least in part to
star-forming activity at a level too weak to dominate the optical continuum
colours.

\begin{figure}

\centerline{\psfig{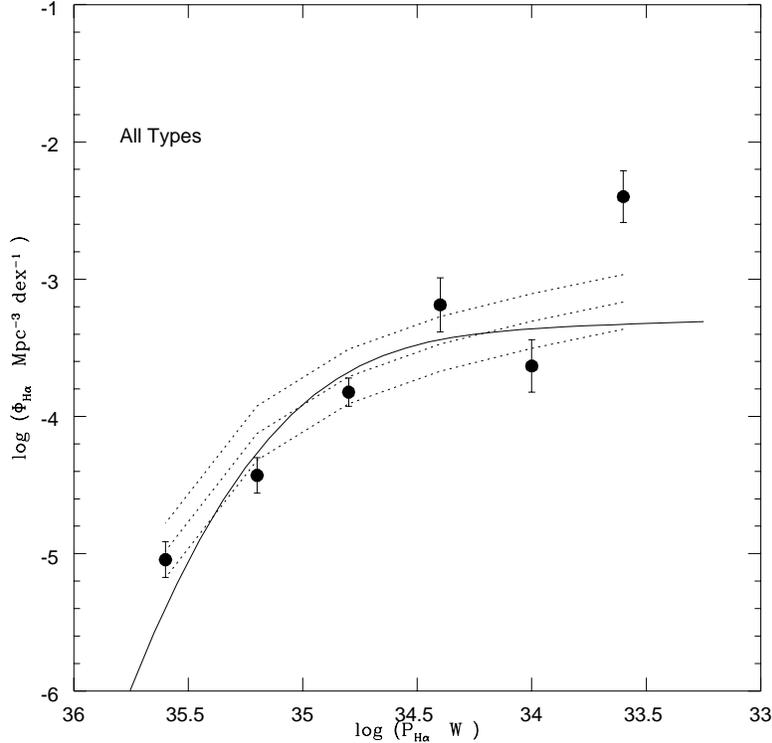}} 
\caption{ The H$\alpha$ luminosity function with a parametric fit 
    (solid lines) for all galaxies with H$\alpha$ detection in the
    radio survey. This is compared with the
    $H\alpha$ luminosity function from an $H\alpha$-selected sample of star
    forming galaxies \protect\cite{Gal:95} - (dotted lines).}
\label{Fig6}
\end{figure}

\begin{table*}
\pagestyle{empty}
\caption{Parameters of the analytical fits to the H$\alpha$
  luminosity functions and the $<V/V_{max}>$ estimates. The results are
  presented for all the faint radio sources in the present survey and for
  the starforming population}
\begin{tabular}{lccccccc}
          &     &                      &      &      &                     \\
  & $n$ & $ <V/V_{max}> $ & $P^\ast_{H\alpha}$ & $\alpha$ & $\sigma$ &
  $C^\ast_{H\alpha}$ \\ 
  & & & W& & & Mpc$^{-3}$\\ 
  & & &  & & & \\ 
All types & 132 &0.47& $(0.45^{+0.03}_{-0.05})\times 10^{35}$ & 
$1.05^{+0.05}_{-0.07}$ & $0.33^{+0.03}_{-0.05}$ & 
$(4.18^{+0.40}_{-0.7})\times 10^{-4}$\\
          &     &                      &      &      &                     \\
Starforming & 75 &$0.45$ & $(0.28^{+0.01}_{-0.03})\times 10^{35}$ & 
$0.89^{+0.02}_{-0.02}$ & $0.40^{+0.01}_{-0.04}$ & 
$(4.62^{+0.40}_{-0.70})\times 10^{-4}$\\
\end{tabular}
\end{table*}

\subsection{Luminosity density distributions and star formation rates}

The main source of uncertainty in converting the observed luminosity to
star formation rate (SFR) is the dependence of the calibration on a number
of poorly known parameters, including the initial mass function and the
history of star formation prior to the epoch of measurement. Differences of
at least a factor of 2 exist between different estimates of SFR due to
these parameters. To overcome this problem, we define the luminosity
density distribution as
\[
{{\cal L}_k (P_k)} = P_k \ \Phi_k (P_k), 
\]
where $k=1, 2$ correspond respectively to 1.4~GHz and H$\alpha$
wavelengths.  This distribution provides an intermediate step between the
luminosity function and the star formation rate density and can be
determined without the uncertainties due to calibration of star formation
rate {\it versus} the luminosity. 

Luminosity density distributions for the radio and H$\alpha$ samples are
presented in Figures 7(a) and 7(b) respectively.  The integral of these
luminosity density distributions give the luminosity densities, ${\cal
  L}_k$, for the radio and H$\alpha$ samples, as listed in Table 2.  The
1.4~GHz luminosity density for the star-forming population in the range $z
\lesssim 1$ ($2.63 \times 10^{19}$ W Hz$^{-1}$ Mpc$^{-3}$) is in excellent
agreement with the value of $2 \times 10^{19}$ W Hz$^{-1}$ Mpc$^{-3}$
determined by Condon \shortcite{Con:89}.  Also, the H$\alpha$ luminosity
density for the faint radio sources in this study ($0.91 \times 10^{32}$ W
Mpc$^{-3}$) is consistent with the value of ${\cal L}_{H\alpha} = 1.26
\times 10^{32}$ W Mpc$^{-3}$ determined by Gallego et al.\ 
\shortcite{Gal:95}, using an H$\alpha$ selected low-dispersion
objective-prism survey of nearby ($z < 0.045$) galaxies.  Considering that
these surveys are entirely independent and have different selection
criteria, the agreement is remarkably good.

\begin{figure} 
\centerline{\psfig{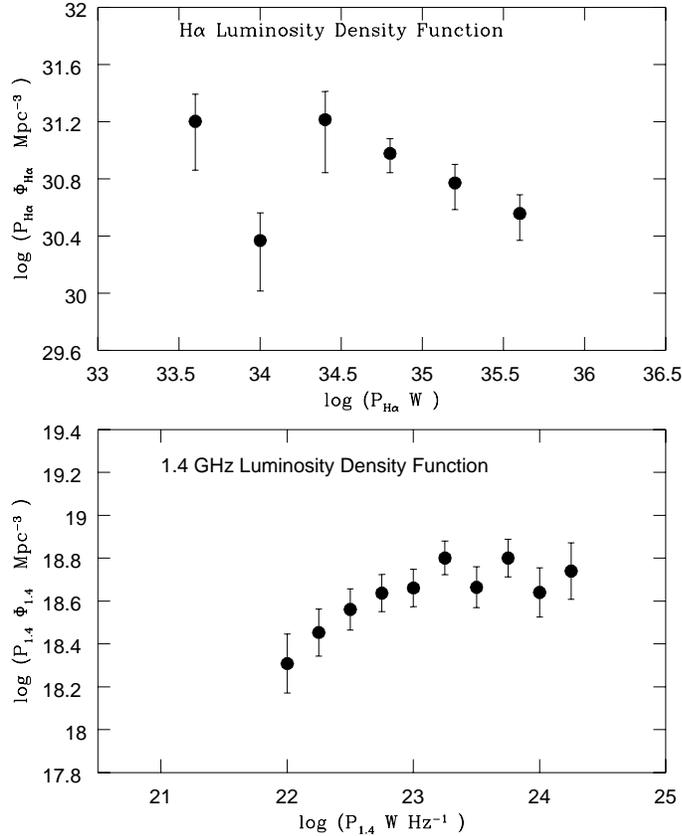}}
\caption{The luminosity density distributions, $P_k\ \Phi (P_k)$, for the
    1.4~GHz radio (a) and H$\alpha$ (b) surveys of faint radio sources. The
    error bars correspond to Poissonian uncertainties. } 
\end{figure}

The relative contribution from the star-forming and elliptical populations
to the radio luminosity densities are shown as a function of their radio
luminosity in Fig. 8.  The star-forming population dominates the radio
luminosity density at $P_{1.4} \simeq 10^{22}-10^{23}$ W Hz$^{-1}$ while
the ellipticals (possibly with AGN activity) contribute significantly at
$P_{1.4} \gtrsim 10^{23}$ W Hz$^{-1}$.  The implication of this result is
that the main contribution to the radio luminosity density at high
redshifts, using flux-limitted radio surveys, is from the elliptical/AGN
population (because these dominate the bright-end of the radio LF)-(Fig.
8).

The luminosity density distribution, ${{\cal L}_k (P_k)}$, can be converted
to the star formation rate density, SFR$_k$, using model-dependent
conversion factors corresponding to
\[
SFR_{1.4} = \frac{{\cal L}_{1.4}}{8.07 \times 10^{20} 
{\mathrm W Hz^{-1}}}~~~{\mathrm
  M_\odot yr^{-1}}
\]
after Condon \shortcite{Con:92} and 
\[
SFR_{\alpha} = \frac{{\cal L}_{H\alpha}}{2.85 \times 10^{33} 
{\mathrm W}}~~~{\mathrm
  M_\odot yr^{-1}}
\]
after Gallego et al.\ \shortcite{Gal:95}. The conversion factors and hence
the resulting SFR densities have been adjusted to an initial mass function
$\psi (M) \propto M ^{-2.35}$ for $0.1 < M / M_\odot < 100$.  Intercomparison
between SFR$_{\alpha}$ and SFR$_{1.4}$ relies on the validity of the
calibration of the 1.4~GHz luminosity of galaxies in terms of their past
supernova rate \cite{Con:90}. As discussed by Cram et al.\ 
\shortcite{Cra:98a} and Cram \shortcite{Cra:98b}, this calibration is quite
uncertain.  Using the above calibrations, the global SFR densities are
estimated and listed in Table 2.  The values of SFR$_{1.4}$ are derived
from the radio luminosity densities of galaxies with optical spectral
morphologies classified as star forming, while the values of SFR$_{\alpha}$
are based on the H$\alpha$ luminosity density for all galaxies with
H$\alpha$ detected.

\begin{table*}[h]
  \pagestyle{empty}
\caption{Integrated luminosity densities and star formation rate
  densities for the faint radio population. The SFR densities are derived
  from both the 1.4~GHz and the H$\alpha$ measurements. The results are
  presented for different types of objects in the sample.}
\begin{tabular}{lccccccc}
  & $z$ & $n_{1.4}$ & $log({\cal L}_{1.4})$ & log ($(SFR)_{1.4}$) & 
$n_{H\alpha}$ &
$log({\cal L}_{H_\alpha})$ & log ($(SFR)_{H_\alpha}$) \\
  &     &     & W Hz$^{-1}$ Mpc$^{-3}$   &M$_\odot$ yr$^{-1}$ Mpc$^{-3}$ 
&     &W  &M$_\odot$ yr$^{-1}$ Mpc$^{-3}$  \\
  &     &     &                   &               &     &  & \\
All Types &     &     &                   &               &     &  & \\ 
  & $0-0.3$ &97 & $19.47^{+0.05}_{-0.06}$ &   & 72 
& $31.91^{+0.12}_{-0.16}$  & $-1.55$\\
  &     &     &                   &               &     &  & \\
   & $0.3-1$ &95 & $19.63^{+0.05}_{-0.06}$  &  & 60 
& $30.96^{+0.08}_{-0.09}$  & $-2.50$\\
  &     &     &                   &               &     &  & \\
   & $0-1$ &192 & $19.86^{+0.04}_{-0.04}$  &   & 132 
& $31.96^{+0.11}_{-0.14}$  & $-1.50$\\
Starforming  &     &     &                   &               &     &  & \\
  & $0-0.3$ &52 & $19.27^{+0.07}_{-0.08}$  & $-1.64$ &  &  & \\
   &     &     &                   &               &     &  & \\
  & $0.3-1$ &23 & $18.88^{+0.09}_{-0.11}$  & $-2.02$ &  &  & \\
  &     &     &                   &               &     &  & \\
  & $0-1$ &75 & $19.42^{+0.06}_{-0.07}$  & $-1.49$ &75  
&$31.92^{+0.11}_{-0.11}$   &$-1.53$ \\
  &     &     &                   &               &     &  & \\
Ellipticals  &     &     &                   &               &     &  & \\
  & $0-0.3$ &26 & $18.79^{+0.08}_{-0.10}$  &  &  &  & \\
  &     &     &                   &               &     &  & \\
   & $0.3-1$ &37 & $19.23^{+0.08}_{-0.10}$  &  &  &  & \\
  &     &     &                   &               &     &  & \\
   & $0-1$ &63 & $19.36^{+0.06}_{-0.08}$  &  &$7$  
&$29.20^{+0.16}_{-0.25}$   &$-4.26$ \\
\end{tabular}
\end{table*}    

\begin{figure}
\centerline{\psfig{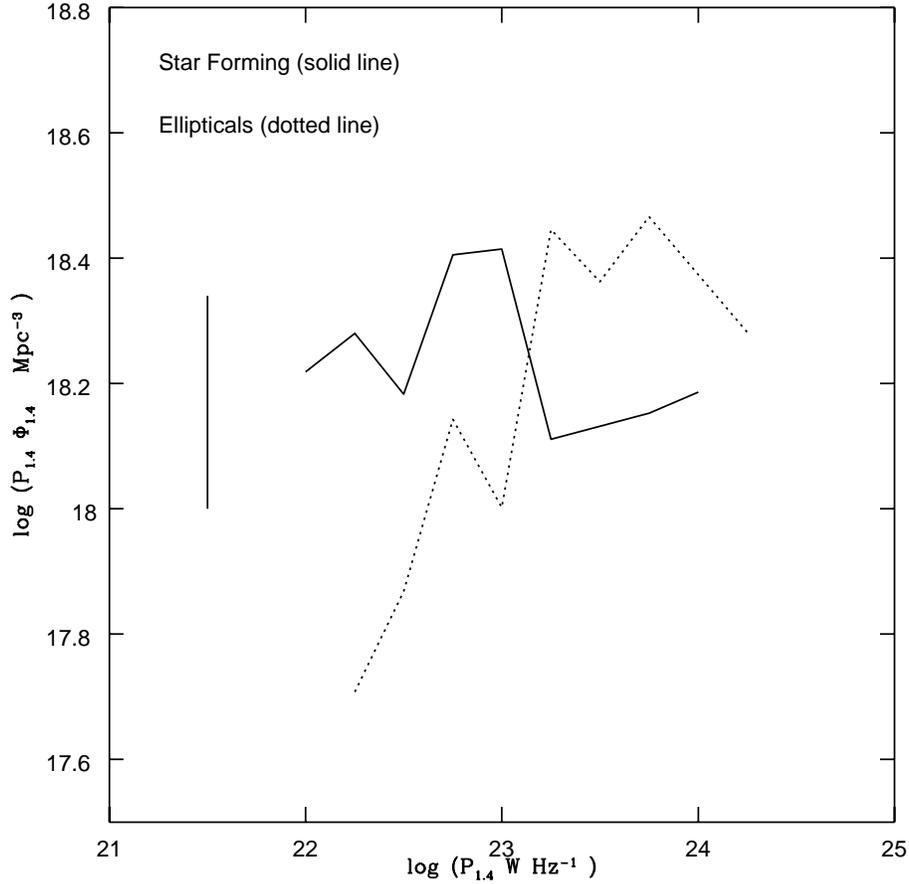}} 
  \caption{The 1.4~GHz luminosity density distributions, 
   $P_{1.4}\ \Phi (P_{1.4})$, for the star forming population (solid line)
   and elliptical galaxies (dotted line). }
\label{Fig8}
\end{figure}
 
We find SFR$_{1.4} = 0.032$ M$_\odot$ yr$^{-1}$ Mpc$^{-3}$ for $z < 1$, is
close to the value estimated from optical studies (Madau et al. 1996; Hogg
et al. 1998).  Also, it is similar to the value SFR$_{\alpha} = 0.031$
M$_\odot$ yr$^{-1}$ Mpc$^{-3}$ we find for galaxies in the same redshift
range.  There is a rather large difference between SFR$_{1.4}$ and
SFR$_\alpha$ for the range $0.3 < z < 1$. This may be due to a combination
of several effects, including small-number statistics, greater extinction
at higher redshift and errors arising from the use of [OII]3727 as a proxy
for $H\alpha$.

The value of SFR$_{\alpha}$ derived from our estimate of the H$\alpha$
luminosity density can be compared directly with that of Gallego et
al. (1995) provided that we ensure that the conversion factors refer to the
same initial mass function. Our value, SFR$_{\alpha} = 0.032$ M$_\odot$
yr$^{-1}$ Mpc$^{-3}$ for $z < 1$ is close to that value of 0.043 M$_\odot$
yr$^{-1}$ Mpc$^{-3}$ found by Gallego et al. (1995) when adjusted to the
same IMF.  Our H$\alpha$ sample is dominated by star-forming (disk)
galaxies, implying that the star-forming radio sources are similar to those
selected at optical (i.e. H$\alpha$) wavelengths.

It is worth noting that the Gallego et al. (1995) sample only covers
galaxies in the range $z < 0.045$. Considering this and the results in
Table 2, we have not been able to establish evidence of cosmic evolution in
the global star formation rate (Table 2), despite the fact that some of the
radio-selected star-forming galaxies lie at redshifts $z > 0.5$ where the
global $SFR$ might exceed the local rate by a factor of the order of
$10^{0.5}$ \cite{Mad:96}.  This conclusion reflects the statistical
uncertainty of our small sample size and is consistent with other similar
investigations (Tresse and Maddox 1998).

\section{Conclusion}

A sample of over 1000 faint radio sources, selected at 1.4~GHz and 
`complete' (to $50\%$) to a flux density limit of $\approx 0.2 $~mJy, 
is used for
this study. Complete $R$-band optical photometry is carried out to $R=22.5$, 
with the optical spectroscopy successfully attempted for over 225 galaxies
selected at random from the radio sources with optical counterparts brighter 
then $R=21.5$. The spectra have been used to derive redshifts and to make 
spectroscopic classifications of the optical galaxies into elliptical, 
star-forming and Seyfert classes. The statistical analysis is carried out on 
the `complete' radio survey and the faint radio sources showing 
H$\alpha$ emissions in their spectra. Conclusions from this study are summarised
as follows:
\begin{enumerate} 
\item Taking into account the various selection effects which define the
  two-wavelength sample, we have derived the radio and H$\alpha$ luminosity
  functions for both starforming and elliptical populations in the interval 
  $0 < z <1$. The LFs are well fitted to models, assuming
  no evolution or luminosity evolution
  in the form $P_{1.4}(z) = P_{1.4}(0) (1+z)^3$. Therefore, the 
  faint radio sources are not affected by luminosity evolution at the
  median redshift ($ z <0.3 $) of the present sample. 

  Both the radio and H$\alpha$ LFs agree closely with their
  counterparts obtained from studies of nearby ($z
  \lesssim 0.05$) galaxies.

\item The radio (1.4~GHz) LF of the star-forming population shows a steeper
  faint-end slope than its counterpart for the ellipticals.  This implies
  that there is a large population of faint radio sources undergoing star
  formation at $z\sim 0.3$, corresponding to the median redshift
  of the present sample. 

\item The luminosity densities are estimated and used to derive the star
  formation rate densities for both the radio and H$\alpha$ data.  These
  give a similar SFR density of 0.032 M$_\odot$ yr$^{-1}$ Mpc$^{-3}$ at
  $z \sim 0.3$. Moreover, The SFR density estimated here is close to those
  found from optical studies. This indicates the efficiency of radio
  observations in selecting the star-forming population, free from biases
  induced by dust.

\item The SFR density based on the H$\alpha$ data, due to the radio sources 
  in the range $0 < z < 1$, is 0.032 M$_\odot$ yr$^{-1}$ Mpc$^{-3}$ 
  which is close to the local 
  value of 0.042 M$_\odot$ yr$^{-1}$ Mpc$^{-3}$ derived by 
  Gallego et al. (1995). Assuming no luminosity evolution in the
  faint radio population, this 
  suggests that the star-forming behaviour of faint radio sources is
  similar to that of star-forming galaxies selected at optical wavelengths.
\end{enumerate}

The Australia Telescope is funded by the Commonwealth of Australia for
operation as a National Facility managed by CSIRO. 
LC and AH acknowledge financial support from the Australian Research Council and
the Science Foundation for Physics within the University of Sydney. We
thank Drs Carole Jackson and Elaine Sadler for several helpful discussions.


\begin{thebibliography}{}


  \bibitem[\protect\citename{Baum \& Heckman\ }1989]{Bau:89}
    Baum S. A., Heckman T., 
    1989, ApJ, 336, 681

  \bibitem[\protect\citename{Benn et al.\ }1993]{Ben:93}
    Benn C. R., Rowan--Robinson M., McMahon R. G., Broadhurst T. J.,
    Lawrence A.,
    1993, MNRAS, 263, 98

  \bibitem[\protect\citename{Blain et al.\ }1998]{Bla:98}
    Blain A. W., Smail I., Ivison R., J., Kneib J.-P.,
    1998, MNRAS (in press)

  \bibitem[\protect\citename{Bruzual \& Charlot\ }1993]{Bru:93}
    Bruzual A. G., Charlot S.,
    1993, ApJ, 405, 538

  \bibitem[\protect\citename{Condon et al.\ }1982]{Con:82}
    Condon J. J., Condon M. A., Gisler G., Puschell J. J., 
    1982, ApJ, 252, 102

  \bibitem[\protect\citename{Condon\ }1984]{Con:84}
    Condon J. J.,
    1984, ApJ, 287, 461

  \bibitem[\protect\citename{Condon\ }1989]{Con:89}
    Condon J. J.,
    1989, ApJ, 338, 13

  \bibitem[\protect\citename{Condon\ }1992]{Con:92}
    Condon J. J.,
    1992, ARA\&A, 30, 575  

  \bibitem[\protect\citename{Condon \& Yin\ }1990]{Con:90}
    Condon J. J., Yin Q. F.,
    1990, ApJ, 357, 97

  \bibitem[\protect\citename{Cram\ }1998]{Cra:98a}
    Cram L. E.,
    1998, ApJ, 506, L85

  \bibitem[\protect\citename{Cram et al.\ }1998]{Cra:98b}
    Cram L. E., Hopkins A. M., Mobasher B. M., Rowan-Robinson M. R.,
    1998, ApJ, 507, 155

  \bibitem[\protect\citename{Danese et al.\ }1987]{Dan:87}
    Danese L., De Zotti G., Franceschini A., Toffolatti L.,
    1987, ApJ, 318, L15

  \bibitem[\protect\citename{Fomalont et al.\ }1984]{Fom:84}
    Fomalont E. B., Kellerman K. I., Wall J. V., Weistrop D.,
    1984, Science, 225, 23

  \bibitem[\protect\citename{Franceschini et al.\ }1988]{Fra:88}
    Franceschini A., Danese L., De Zotti G., Toffolatti L.,
    1988, MNRAS, 233, 157

  \bibitem[\protect\citename{Franceschini et al.\ }1994]{Fra:94}
    Franceschini A., Mazzei P., De Zotti G., Danese L.,
    1994, ApJ, 427, 140

  \bibitem[\protect\citename{Gallego et al.\ }1995]{Gal:95}
    Gallego J., Zamamoro J., Arag\'{o}n-Salamanca A., Rego M.,
    1995, ApJ, 455, L1

  \bibitem[\protect\citename{Geogakakis et al.\ }1998]{Geo:98}
    Georgakakis A., Mobasher, B., Hopkins A., Cram L., Lidman C.,
    Rowan-Robinson M.,
    1998, MNRAS accepted astro-ph 9903016

  \bibitem[\protect\citename{Hammer et al.\ }1995]{Ham:95}
    Hammer F., Crampton D., Lilly S., le F\`{e}vre O., Kenet T.,
    1995, MNRAS, 276, 1085

  \bibitem[\protect\citename{Hine \& Longair\ }1979]{Hin:79}
    Hine R. G., Longair M. S.,
    1979, MNRAS, 188, 111

  \bibitem[\protect\citename{Hopkins et al.\ }1998]{Hop:98}
    Hopkins A. M., Mobasher B., Cram L., Rowan-Robinson M.,
    1998a, MNRAS, 296, 839

  \bibitem[\protect\citename{Hopkins et al.\ }1999]{Hop:99}
    Hopkins A., Cram L., Mobasher B., Georgakakis A. 
    1999, in ``Looking Deep in the South'' (in press)

  \bibitem[\protect\citename{Kauffmann \& Charlot\ }1998]{Kau:98}
    Kauffmann G., Charlot S.
    1998, MNRAS, 297, L23

  \bibitem[\protect\citename{Kawara et al.\ }1998]{Kaw:98}
    Kawara et al. 1998, A \& A 336, L9.

  \bibitem[\protect\citename{Kennicutt\ }1992]{Ken:92}
    Kennicutt, R. C.,
    1992, ApJ, 388, 310

  \bibitem[\protect\citename{Kron et al.\ }1985]{Kro:85}
    Kron R. G., Koo D. C., Windhorst R. A.,
    1985, A\&A, 146, 38

  \bibitem[\protect\citename{Lilly et al.\ }1996]{Lil:85}
    Lilly, S. J., Le Fevre, O., Hammer, F., Crampton D., 1996, Ap.J. 460, L1

  \bibitem[\protect\citename{Lonsdale et al.\ }1987]{Lon:90}
    Lonsdale C. L., Hacking P. B., Conrow T. P., Rowan-Robinson M.,
    1990, ApJ, 358, 60 

  \bibitem[\protect\citename{Madau et al.\ }1996]{Mad:96}
    Madau P., Ferguson H. C., Dickinson M. E., Giavalisco M., Steidel
    C. C., Fruchter A.,
    1996, MNRAS, 283, 1388

  \bibitem[\protect\citename{Mitchell \& Condon\ }1985]{Mit:85}
    Mitchell K. J., Condon J. J., 
    1985, AJ, 90, 1957

  \bibitem[\protect\citename{Rowan-Robinson et al.\ }1993]{Row:93}
    Rowan-Robinson M., Benn C. R., Lawrence A., McMahon R. G., Broadhurst
    T. J.,
    1993, MNRAS, 263, 123

  \bibitem[\protect\citename{Rowan-Robinson et al.\ }1997]{Row:97}
    Rowan-Robinson M. R., et al.
    1997, MNRAS, 289, 490

  \bibitem[\protect\citename{Sadler et al.\ }1989]{Sad:89}
    Sadler E., Jenkins C., Kotanyi C.,
    1989, MNRAS, 240, 591

  \bibitem[\protect\citename{Sanders \& Mirabel\ }1996]{San:96}
    Sanders D. B., Mirabel I. F.,
    1996, ARA\&A, 34, 749

  \bibitem[\protect\citename{Saunders et al.\ }1990]{Sau:90}
    Saunders W., Rowan-Robinson M., Lawrence A., Efstathiou G., Kaiser N.,
    Ellis R. S., Frenk C. S.,
    1990, MNRAS, 242, 318

  \bibitem[\protect\citename{Thaun \& Condon\ }1987]{Thu:87}
    Thuan T. X., Condon J. J., 
    1987, ApJ, 322, L9
 
  \bibitem[\protect\citename{Tresse \& Maddox\ }1998]{Tre:98}
    Tresse L., Maddox S. J.,
    1998, ApJ, 495, 691

  \bibitem[\protect\citename{Wall \& Jackson\ }1997]{Wal:97}
    Wall J. V., Jackson C. A., 
    1997, MNRAS, 290, L17

  \bibitem[\protect\citename{Windhorst et al.\ }1984]{Win:84}
    Windhorst R. A., van Heerde G. M., Katgert P., 
    1984, A\&AS, 58, 1 

  \bibitem[\protect\citename{Windhorst et al.\ }1985]{Win:85}
    Windhorst R. A., Miley G. K., Owen F. N., Kron R. G., Koo D. C.,
    1985, ApJ, 289, 494

\end{thebibliography}
\end{document}